\documentclass[runningheads]{llncs}

\usepackage{graphicx}
\usepackage{booktabs}
\usepackage{rotating}
\usepackage{ragged2e}
\usepackage{setspace}
\usepackage{amssymb}
\usepackage{footnote}
\usepackage{threeparttable}
\usepackage{vwcol}  
\usepackage{chngpage}
\usepackage{todonotes}
\usepackage{float}

\begin{document}
\title{Test-case quality -- understanding practitioners' perspectives 
}

\author{Huynh Khanh Vi Tran\inst{1} \and \\
Nauman bin Ali\inst{2} \and
J\"urgen B\"orstler\inst{3} \and
Michael Unterkalmsteiner\inst{4}}

\authorrunning{V. Tran et al.}
\institute{SERL Sweden, Blekinge Institute of Technology, SE-371 79 Karlskrona, Sweden \\
\email{\{\inst{1}huynh.khanh.vi.tran, \inst{2}nauman.ali, \inst{3}jurgen.borstler, \\ \inst{4}michael.unterkalmsteiner\}@bth.se}}
\maketitle           
\begin{abstract}

Background: Test-case quality has always been one of the major concerns in software testing. To improve test-case quality, it is important to better understand how practitioners perceive the quality of test-cases.

Objective: Motivated by that need, we investigated how practitioners define test-case quality and which aspects of test-cases are important for quality assessment.

Method: We conducted semi-structured interviews with professional developers, testers and test architects from a multinational software company in Sweden. Before the interviews, we asked participants for actual test cases (written in natural language) that they perceive as good, normal, and bad respectively together with rationales for their assessment. We also compared their opinions on shared test cases and contrasted their views with the relevant literature.

Results: We present a quality model which consists of 11 test-case quality attributes. We also identify a misalignment in defining test-case quality among practitioners and between academia and industry, along with suggestions for improving test-case quality in industry.

Conclusion: The results show that practitioners' background, including roles and working experience, are critical dimensions of how test-case quality is defined and assessed.

\end{abstract}

\section{Introduction}
Testing plays an important role in software quality assurance, which has been one of the main concerns in the software development life cycle.
The fundamental artefacts in testing are test cases. 
Grano et al. have shown in their study that good test cases in terms of being simple and readable make it easier for developers to maintain them and to keep up with fast software development life cycle~\cite{grano2018empirical}.
A study by Athanasiou et al. also showed that high quality of test code could also increase the performance of development teams in fixing bugs and implementing new features~\cite{athanasiou2014test}.
Therefore, good test cases increase the confidence in testing, and thereby assist product release decisions.
Hence, assuring the quality of test cases is an important task in quality assuring software-intensive products.

There have been studies which focused on different test-case quality attributes such as performance, readability, and effectiveness~\cite{causevic2012test,garousi2016developing,gopinath2014code,grano2018empirical,inozemtseva2014coverage,nagappan2005providing,pfaller2008multi,reichhart2007rule,zhu1997software}.
Some studies adapted the ISO standard for software quality to define test-case quality~\cite{neukirchen2008approach,zeiss2007applying}.
Those studies provided researchers' perceptions of test-case quality.
Though the contributions from academia are important, it is necessary to verify how knowledge could be transferred between academia and industry.
The first step would be to investigate how test-case quality is understood by practitioners.
However, there is currently a lack of empirical studies on the topic.

To reduce this gap, we conducted an exploratory study to investigate how test-case quality is defined and assessed in practice.
Our focus was manual test cases written in natural language.
This type of test cases is still required for testing levels such as system testing, acceptance testing, and for a testing approach such as exploratory test.
Hence, studying how the quality of natural-language tests is perceived in practice is as important as of code-based test cases.
The contributions of the study are as follows:
\begin{itemize}
    \item Descriptions of test-case quality attributes identified by practitioners.
    \item Reasons for the difference in defining and assessing test-case quality among practitioners with different roles, and between academia and practice.
    \item Context factors to consider when defining test-case quality.
    \item Suggestions to improve test-case quality by practitioners.
    \item Sources of information for understanding and assessing test-case quality suggested by practitioners.
\end{itemize}

The remainder of the paper is structured as follows: Section~\ref{sec:relatedWork} describes related work. 
Section~\ref{sec:researchMethod} describes the study design, followed by Section~\ref{sec:validityThreats} which discusses threats to validity. Section~\ref{sec:results} discusses our findings. Our conclusions and future work are summarised in Section~\ref{sec:results}.

\section{Related Work}\label{sec:relatedWork}
We identified nine studies which involved practitioners in their work focusing on test-case quality~\cite{athanasiou2014test,bowes2017good,chernak2001validating,daka2015modeling,garousi2018we,garousi2018smells,hauptmann2013hunting,jovanovikj2018context,adlemo2018test}. 
We organised them into three groups.

The first group includes two studies which integrated practitioners' knowledge into the studies' results regarding test-case quality~\cite{adlemo2018test,bowes2017good}.
Adlemo et al.~\cite{adlemo2018test} introduced 15 criteria for good test cases.
There was no specific focus on types of test cases.
Of those criteria, ten were inspired by the literature while five came from practitioners' suggestions. 
The criteria were ranked by 13 Swedish practitioners with experience in software testing and software development.
\textit{Repeatability}, meaning that a test case should produce the same result whenever it receives the same input, had the highest votes from practitioners.
Bowes et al.~\cite{bowes2017good}, focused on test code in unit testing.
The authors identified 15 testing principles collected from three sources: a workshop with industrial partners, their software testing teaching materials, and practitioners' books.
\textit{Simplicity} in terms of test-code size, number of assertions and conditional test logic, is considered as the most important principle, and is the foundation for the other ones.

The second group contains four studies which had practitioners evaluate their hypotheses relating to some test-case quality attributes~\cite{athanasiou2014test,chernak2001validating,daka2015modeling,jovanovikj2018context}.
Jovanovikj et al.~\cite{jovanovikj2018context} introduced an approach and a tool to evaluate and monitor test-case quality.
They presented eight quality characteristics based on Zeiss et al.'s work~\cite{zeiss2007applying}, which relied on the ISO/IEC 25010 (ISO/IEC, 2011) software quality models.
To verify their approach's applicability, they conducted a case study in the context of natural-language tests, and had interviews with two quality managers and some testers.
Similarly, Athanasiou et al.~\cite{athanasiou2014test} proposed a model to assess three test-code quality attributes, namely Completeness, Effectiveness, and Maintainability with associated metrics.
To verify if the model was aligned with practitioners' opinions, they compared its results from two software systems with the evaluations of two experts via focused interviews.
They concluded that there is a strong correlation between test code quality, throughput, and productivity of issue handling.
In another study, Daka et al.~\cite{daka2015modeling} introduced a model of unit test readability which could help to generate test suites with high coverage and high readability.
Their model involved human judgement, but there was no clear indication on their selection criteria.
Focusing on only test-case effectiveness, Chernak~\cite{chernak2001validating} proposed an evaluation and improvement process for test cases in general.
The process was used by one project team, including three testers and 10 developers who worked on a banking system.

The third group includes three studies which discussed test smells~\cite{garousi2018we,garousi2018smells,hauptmann2013hunting}.
Hauptman et al.~\cite{hauptmann2013hunting} presented seven test smells in natural-language tests, which were collected based on their experiences with evaluating natural-language system tests.
Their study was claimed as the first work on test smells in the context of natural-language tests.
For smells in test code, Garousi et al.~\cite{garousi2018smells} conducted a systematic ‘multivocal’ literature mapping and developed a classification for 196 test smells.
The authors included their descriptions of top 11 most discussed test smells in a subsequent study~\cite{garousi2018we}.

The related works show that practitioners' perceptions of test-case quality have not been well studied.
Particularly, we have not identified any study focusing on eliciting first-hand data from practitioners on their perceptions of test-case quality in the context of natural language tests.

\section{Research Method}\label{sec:researchMethod}
The objective of the study is to gain a better understanding of practitioners’ perceptions towards test-case quality.
We conducted an exploratory study with a multinational telecommunication company in Sweden. 
This type of study was chosen since the research focus on eliciting practitioners’ genuine perspective on test-case quality has not been well studied.
The exploratory study helped us to get more familiar with the research topic, to determine what the study design(s) should be for our subsequent studies on the same topic. 

In this study, we used semi-structured interviews to explore the practitioners' perspectives on the topic.
According to Robson and McCartan~\cite{robson2016real}, this interview approach allows researchers to flexibly modify the interview questions depending on the interviewees’ answers. 
Since the interviews were about discussing test-case quality, the same strict questionnaire would not be applicable to all interviewees.
Also, the interviews were based on real test cases provided by the interviewees.
Thanks to the explicit test cases, our approach makes it easier for interviewees to refer to instances of quality aspects instead of vague, generic or abstract ideas.

\subsection{Research Questions}\label{sec:RQs}
\begin{itemize}
    \item \textit{RQ1. How do practitioners describe test-case quality?} 
    The research question directly connects to our study's objective. 
    Without defining quality criteria upfront, we first want to elicit information on how practitioners perceive test-case quality.
    \item \textit{RQ2. How well is the understanding of test-case quality aligned among practitioners in a company?}
    Test-case quality might be assessed differently depending on how it is perceived by the assessors.
    That could affect testing-related activities such as test-case design, and test-case maintenance.
    Therefore, we want to understand whether practitioners perceive test-case quality differently; if so, then we want to identify the potential reasons.
    \item \textit{RQ3. What context factors do practitioners consider when assessing test-case quality?}
    The context factors could be testing level, testing purpose, characteristics of the software system under test, etc.
    Test-case quality might be context-dependent.
    Hence, we want to identify the potential context factors or aspects which could influence how test-case quality is assessed.
    \item \textit{RQ4. What are potential improvements suggested by practitioners for improving test-case quality?}
    Answers to this research question would help us to understand practitioners' needs regarding test-case quality, which could give us and researchers potential research directions.
    \item \textit{RQ5. What information sources guide practitioners' opinions of test-case quality?}
    Identifying such information sources could helps us to understand why practitioners perceive test-case quality in certain ways.
\end{itemize}

\subsection{Data Collection}
The data was collected from the interviews which included test cases provided by the interviewees.
Before conducting the study, we had a meeting with the company's representatives to present our study's design and to obtain basic understanding of the company's structure, and potential interviewees.

\subsubsection{Interview Design}
Before conducting the interviews, we asked each participant to provide us three test cases with their quality classification (good, bad or normal).
They could choose any test case from the company's test suites that they are familiar with. 
We also asked for a written rationale for the classification, since we not only wanted to see whether other interviewees would rate them similarly, but also whether they would provide similar reasons. 
We intentionally did not define quality criteria upfront in order to elicit the genuine perceptions of the interviewees.
We swapped the test cases between two participants who work in the same team.
Before the interviews, we informed the interviewees which test cases they had to review extra.
Hence, in the interviews, the swapped test cases were also judged by the interviewees so that we could gauge their alignment.

We used the pyramid model~\cite{robson2016real} for our interview session.
Hence, each interview starts with specific questions followed by more open questions.
More specifically, the interview session is divided into three phases. 
\begin{itemize}
    \item Part 1: Background Information: we focused on obtaining information about the interviewee’s testing experience.
    \item Part 2: Test Case Classification: we asked the interviewee to clarify his reasons for his test-case quality classification and to discuss some test cases given by another participant.
    \item Part 3: General Perception of Test-Case Quality: we had a more generic discussion with the interviewee about his or her perception of test-case quality.
\end{itemize}

To mitigate flaws in our interview design, we conducted a pilot interview with a colleague whose research interest includes software testing and has been working with test cases for years. 
The interview questionnaire could be found at https://tinyurl.com/y6qakcjc.

\subsubsection{Participants Selection}

Our selection criteria were that (1) a participant should be a tester and/or a developer; (2) the participant has at least one year of working experience relating to software testing.
Our selection is convenience sampling~\cite{kitchenham2002principles} as we involved those who meet our criteria and are willing to participate in the interviews.
At the end, we had six participants from three different teams working in different projects. 
Their information is described in Table~\ref{tab:participantInfo}. 
Even though four of them are test architects, their responsibilities still involve working with test cases. 
Hence, having them participate in the study did not affect our study design.

\begin{table}[!ht] 
{
    \RaggedRight
    \footnotesize
    \begin{center}
    \caption{Participants' Experience, Roles, Tasks and Test Cases Provided}
    \label{tab:participantInfo}
    \begin{threeparttable}
    \begin{tabular}{p{0.035\linewidth}p{0.155\linewidth}p{0.06\linewidth}p{0.072\linewidth}p{0.09\linewidth}p{0.095\linewidth}p{0.095\linewidth}p{0.121\linewidth}p{0.11\linewidth}p{0.08\linewidth}}
        \toprule
        \textbf{ID} & \textbf{Role} & \textbf{Exp}\tnote{1} & \textbf{Make TP}\tnote{2} & \textbf{Design TCs}\tnote{3} & \textbf{Review TCs} & \textbf{Report TR}\tnote{4} & \textbf{Maintain TCs} & \textbf{Execute TCs}& \textbf{TC ID} \\
        \midrule
        P1 & test \mbox{architect} & 6 & \checkmark & \checkmark & \checkmark & \checkmark & \checkmark & & P1.1-3 \\ \hline
        P2 & tester & 14 & & & & & \checkmark & \checkmark & P2.1-4 \\ \hline
        P3 & test \mbox{architect} & 6 & \checkmark & \checkmark & \checkmark & \checkmark & \checkmark & & P3.1-2 \\ \hline
        P4 & tester, test \mbox{architect}, consultant & 20 & \checkmark & \checkmark & \checkmark & \checkmark & & \checkmark & P4.1-3 \\ \hline
        P5 & developer & 5 & & \checkmark & & & & \checkmark & P5.1-2 \\ \hline
        P6 & test \mbox{architect} & 15 & \checkmark & \checkmark & \checkmark & \checkmark & \checkmark & & P6.1-3 \\
        \bottomrule
    \end{tabular}
    \begin{tablenotes}\footnotesize
    \noindent
    \begin{minipage}[c]{0.7\linewidth}
    \item[1] Exp: number of years of working experience in testing
    \item[2] TP: test plan
    \end{minipage} 
    \begin{minipage}[c]{0.2\linewidth}
    \item[3] TC: test case
    \item[4] TR: test results
    \end{minipage}
    \end{tablenotes}
    \end{threeparttable}
    \end{center}
}
\end{table}

\subsubsection{Interview Execution}
The interviews were conducted by two researchers each. 
One researcher asked questions while the other took notes and added extra questions for clarification if needed. 
Each interview took around one hour, and was audio-recorded with the participant's consent.

\subsubsection{Test Cases}
In total, we collected 17 manual natural-language test cases as not all practitioners followed the instruction of providing three test cases each.
They were extracted from the company’s test suites for functional testing.
We focused on the following information of a test case in our analysis: ID, name, description, and steps.
Even though there is no strict format for the test case's description, it often includes, but does not require, the following information: purpose, preconditions, additional information, references, and revision history.
Additionally, we also received the quality classification (Good/Bad/Normal) and the written explanations before the interviews.
Nonetheless, we could not report the actual test cases' content due to confidentiality reasons.

\subsection{Data Analysis}
\subsubsection{Interview Data}
Before analysing the data, the first author transcribed and anonymised all audio recordings of the interviews.
The transcribed data were coded using a thematic coding approach~\cite{cruzes2011recommended}.
More specifically, we applied an integrated approach, which allows codes to be developed both inductively from the transcribed data and deductively from the research questions and researchers' understanding of test-case quality in general.
The main themes which were inspired by the research questions are as follows:
\begin{itemize}
    \item Practitioners' Background Information: contains information such as roles, testing experience;
    \item Test-Case Quality Description: contains information about how practitioners described test-case quality and their selection of the top three quality indicators or attributes of a good test case and of a bad one;
    \item Test-Case Quality Assessment: contains information about practitioners' classification of test-case quality and their reasoning;
    \item Test-Case Quality Alignment: contains information about differences and similarities in practitioners' perceptions of test cases and their reasoning;
    \item Test-Case Quality Improvement: contains information about practitioners' suggestions to improve test-case quality;
    \item Source of Information: contains information about sources that practitioners refer to when they need to assess or get a better understanding of test-case quality.
\end{itemize}

For each interview, we followed the following steps:
\begin{itemize}
    \item[]Step 1: Starting from the beginning of the interview, mark a segment of text which is relevant to the pre-defined themes with a code and assign it to a corresponding theme.
    For the Test-Case Quality Description theme, relevant codes could be test-case quality attributes such as \textit{understandability}, \textit{effectiveness}, \textit{traceability}, etc.
    Some of those attributes were named and explained explicitly by the practitioners while the others were generated based on their discussions during the interviews.
    \item []Step 2: Find the next relevant segment of text. Mark it with an existing code or with a new code and assign it to a relevant main theme. 
    If the information is related to test-case quality but does not belong to any main theme then a new theme is created for that new information. It helps us to capture emerging concepts related to our study's focus.
    \item []Step 3: Repeat Step 2 until no relevant information is found.
    \item []Step 4: Create sub-themes under every main theme to cluster related codes together.
\end{itemize}

During the process, codes, themes, and their descriptions were continuously refined to fit the data.
We used a commercial tool to complete this coding process, which allows us to maintain traceability between the transcribed data and the related codes and themes.
To mitigate bias and increase the reliability of the coding, the first set of codes and themes were discussed by two researchers, and the coding scheme was refined. 
Furthermore, the final set was reviewed by all researchers. 
All disagreements regarding the coding were resolved in a meeting by discussion.

To obtain an overall ranking of the top quality indicators and attributes of a good test case and of a bad one, each of them gets three points if it was ranked first by a practitioner, two points if it was ranked second, and one point otherwise.
We wanted to get a general picture of which quality attributes or indicators are normally considered more important than the other by practitioners.
Hence, we did not consider the contextual factors identified by RQ3 in the ranking.

\subsubsection{Test Case Data}\label{sec:dataAnalysis_TC}
To analyse the collected test cases, we extracted the quality classifications and reasons from practitioners' written notes.
The information was coded in the same manner as the interview data (see previous section).
To compare practitioners' opinions with the literature, before the interviews, we searched for test smells in those test cases based on test smells' descriptions from two studies~\cite{garousi2018we,hauptmann2013hunting}.
This step did not only give us another assessment angle but also helped us to better understand the test cases' quality.
We selected those studies for reference for two reasons.
The first study~\cite{hauptmann2013hunting} is the most recent work on test smells of natural-language tests.
The second study~\cite{garousi2018we} provides us descriptions of the top 11 most discussed smells of test code.
There are common characteristics between natural language test cases and unit test cases such as testing logic, issues in test steps, dependencies between test cases, test behaviour when executing, etc.
Hence, the study of Garousi et al.~\cite{garousi2018we} is a relevant reference.
Even though that study was based on a former work of Garousi et al.~\cite{garousi2018smells}, the former one did not provide definitions of test smells, hence not chosen as a reference.

\section{Threats to Validity}\label{sec:validityThreats}
\subsubsection{Construct Validity}
Construct validity is concerned with the reliability of operational measures to answer the research questions.
Our interviews were semi-structured with follow-up questions which gave us opportunities to clarify practitioners' answers and reduce misunderstandings during the interview.
Their written explanations for the test cases' quality assessment reduced the risk of misinterpreting their answers.
The test cases were selected subjectively by the practitioners to demonstrate their perspective of good/bad/normal test cases in terms of their perceived quality. 
Since our study's type is exploratory and attempts to capture practitioners perspective, this selection method is not considered a threat to the validity of our results.
Additional information about practitioners such as whether they were ISTQB\footnote{\url{https://www.istqb.org/}}-certified might influence their perspective on test-case quality. Since we did not collect this information, it is a limitation of the study. Nevertheless, we collected important information (their testing experience, roles, and working tasks relating to test cases) which would be still sufficient to describe the participants’ background information.

\subsubsection{Internal Validity}
Internal validity is about causal relations examined in the study. 
Even though we identified possible aspects which should be considered when defining and assessing test-case quality, our focus was not to generate a complete list of such aspects.
By not eliminating one aspect or another, this type of threat is not of concern.

\subsubsection{External Validity}
External validity is concerned with the generalisability of the study's findings.
In general, with the "convenience sampling"~\cite{kitchenham2002principles}, the sample might not represent the population, which could potentially affect the findings' generalisability. 
However, as our study is exploratory, not confirmatory, this sampling method is not considered as a validity threat.
Our study's context is characterised by the type of the company, which is a global company working on embedded software systems, the practitioners’ documented working background and the nature of the natural-language tests.
That is the context to which the findings can be potentially applied.

\subsubsection{Reliability}
Reliability is about the reliability of the results.
Our study's design was discussed among all authors of the paper.
The interviews were conducted by two researchers and the findings were discussed by all researchers to mitigate the bias from one researcher.
The data collection process and interview questions were clearly documented to increase the reliability.

\section{Results and Discussion}\label{sec:results}
In this section, we present and discuss our findings in relation to each of the research questions stated in Section~\ref{sec:RQs}.

\subsection{Test-Case Quality Definition (RQ1)}
To answer the first research question, we asked practitioners to define test-case quality and explain how they would assess such quality (the interview question Q7-11). 
Table~\ref{tab:TCQualityAttribute} contains a list of 11 test-case quality attributes that we collected.
It also includes the practitioners' authentic terms and phrases used to describe the attributes. 
It is worth mentioning that the use of specific test cases, chosen by the participants from the organization's test suites, triggered more in-depth reflections. 
The insights from practitioners regarding these test cases identified as many unique test-case quality attributes as a discussion in abstract of what constitutes test-case quality. 

Overall, we could see that the quality attributes could be placed into two groups.
The first group, including \textit{understandability}, \textit{step cohesion}, \textit{completeness}, \textit{efficiency}, and \textit{flexibility}, is oriented around quality attributes of a test case which could be relevant for practitioners when executing it.
The second group includes \textit{understandability}, \textit{simplicity}, \textit{completeness}, \textit{homogeneity}, \textit{issue-identifying capability}, \textit{repeatability}, \textit{traceability}, \textit{effectiveness}, and \textit{reasonable size}. 
The latter group of attributes relates to general concerns, namely the design, the maintenance, and the objective of testing in general.

\textit{Understandability} is the most common attribute, and discussed by all practitioners. 
A reason for this could be the nature of the discussed test cases, which were written in natural language.
Hence, it makes sense that ambiguity in test cases is considered as a major concern.
We could also see an alignment between practitioners' perceptions and the literature. 
\textit{Understandability} is directly connected to three test smells, namely \textit{ambiguous tests} in natural-language tests~\cite{hauptmann2013hunting}, \textit{long/complex/verbose/obscure test}, and \textit{bad naming} in test code~\cite{garousi2018we}.
Even though the last two smells are for test code, according to their definitions, which are "It is difficult to understand the test at a glance. The test usually has excessive lines of code" and "The test method name is not meaningful and thus leads to low understandability" respectively, those smells could also occur in natural-language tests.
The other connection is between the quality attribute \textit{simplicity} and the test smell \textit{eager test}, which is described as "The test is verifying too much functionality in a single test method"~\cite{garousi2018we}.

Apart from identifying test-case quality attributes, practitioners also listed the top characteristics and indicators of a good test case and of a bad one.
The outcome is a mixture of specific quality indicators: \textit{clear objective} (the purpose of a test case), \textit{clear precondition} (how to set up the testing environment), \textit{clear steps with clear expected results}, and general quality attributes: \textit{understandability}, \textit{completeness}, \textit{effectiveness}.
According to our ranking scheme, \textit{understandability} is rated as the most important attribute.
This is consistent with the most commonly discussed quality attributes in the general discussion.
The second place goes to the quality indicator \textit{clear objective}.
One of the reasons given by one practitioner was that "the objective of each test case or of each component of the test scope is the most important thing because those are combined to make sure that all the requirements of each of the projects are met."

\begin{table}[H]
{
    \setlength{\tabcolsep}{0.35em}
    \footnotesize
    \centering
    \begin{center}
    \caption{Test-Case Quality Attributes}
    \label{tab:TCQualityAttribute}
    \begin{threeparttable}
    \begin{tabular}{p{0.19\linewidth}p{0.37\linewidth}p{0.35\linewidth}p{0.025\linewidth}}
        \toprule
        \textbf{Quality \mbox{Attribute}} & \textbf{Description} & \textbf{Practitioners' phrases}& \textbf{N\tnote{1}} \\
        \midrule
        Understand-ability & The information of a test case (name, objective, precondition, steps, terms) should be easy to understand by both testers, and developers & straightforward, understandable description, how and what to test, clear steps, clear objective, clear precondition & 6 \\
        Simplicity & A test case should not combine different test cases together nor contain so many steps & a big story for many test, not so many steps cases & 4\\
        Step cohesion & Steps in a test case should be well connected. The test case should not contain redundant steps or miss necessary steps & unnecessary step, mandatory steps & 3\\
        
        Completeness & A test case should contain all relevant information for its execution & all information needed to perform the test, all kind of \mbox{information} that developers and testers need & 2\\
        Homogeneity & Test-case design should follow the same rules & homogeneous, unity with the same rules, harmony & 2\\
        Issue-identifying capability & A test case should help to identify issues, weakness of features/functions & find bug, mitigate possible \mbox{issues} & 2\\
        Repeatability & A test case returns the same results every time it is executed &  run any time, tested \mbox{repeatedly} & 2\\ 
        Traceability & There should be traces between a test case and other related artefacts such as issues, ISO quality attributes, functionality & mentioned issue, function \mbox{category}, ISO attributes category & 2\\
        
        Effectiveness & A test case covers the expected requirements & meets the requirement & 1\\
        Efficiency & A test case should be easy to run so that it does not waste time & efficient, easy to run, not complicated, save time & 1\\
        Flexibility & A test case should have flexibility in how to execute it &  flexible, loosely written test, freedom, run differently & 1\\
        
        \bottomrule
    \end{tabular}
    \begin{tablenotes}
    \footnotesize
    \noindent
    \begin{minipage}[c]{1\linewidth}
    \item[1] N: Number of practitioners discussed the quality attribute
    \end{minipage} 
    \end{tablenotes}
    \end{threeparttable}
    \end{center}
}
\end{table}

\subsection{Alignment in Understanding of Test-case Quality (RQ2)} 
We asked practitioners to classify test cases given by the others into \textit{good}, \textit{bad} or \textit{normal} in terms of their quality (Section 2 of the interview questionnaire).
Due to the interviews' time constraint, only seven out of 17 test cases, were analysed by more than one practitioners as shown in Table~\ref{tab:TCQualityClassification}.
Half of them, P1.3, P2.4, P3.1, and P3.2, had the same quality classification while the other half, P1.2, P3.1, P4.1, and P5.2, received a mixed assessment.

In general, we could see that test-case \textit{understandability} was always the first concern. 
For the test cases having the same quality assessment (P1.3, P2.4, P3.1, P3.2), a test case's quality is considered as absolutely bad if the practitioners could not understand what they are supposed to do, especially when both the test objective and other details like steps, precondition, expected results of steps are unclear.
If the test case's objective is sufficiently clear enough that the practitioners could get some idea about its purpose, they would consider its quality as acceptable or normal, though other details like preconditions are missing.

By analysing test cases which had different quality classification results (P1.2, P3.1, P4.1, and P5.2), we could see that the difference is strongly associated with the practitioners' responsibilities relating to test cases.
If one of their responsibilities is to execute test cases, then they are more concerned about whether they have all relevant information to run the test cases.
If they are responsible for broader tasks, in this case mainly about test-case maintenance and test results analysis such as what faults to fix, then they would have other concerns such as the test cases' complexity or their traceability to issues, bugs.

Our observation aligns with the perceptions of practitioners as they explained that they might have different concerns regarding test cases depending on their responsibilities.
Those responsible for executing test cases prioritise \textit{understandability} and \textit{completeness} of test cases, that is, whether they have all relevant and clear information for executing the test cases.
Those responsible for broader tasks like test architects do not only care about how test cases execute but also about the outcome of the test cases and the test suites in general.
Hence, they have extra expectations such as whether the test cases cover the requirements, or whether it is easy to maintain the test cases.
They also explained that the difference in working styles might have an impact on the test-case quality assessment.
If they have different approaches in designing test cases, they would have different requirements on how to assure the test-case quality.

To provide a different perspective on the test-case quality assessment, the lead author used the list of test smells from the literature (see Section~\ref{sec:dataAnalysis_TC}) to identify test smells in those seven test cases.
As a result, there is a considerable overlap between the practitioners' concerns and the identified test smells (\textit{ambiguous test}~\cite{hauptmann2013hunting}, \textit{conditional tests}~\cite{hauptmann2013hunting}, \textit{long/complex/verbose/obscure test}~\cite{garousi2018we}, and \textit{eager test}~\cite{garousi2018we}) (shown in Table~\ref{tab:TCQualityClassification}).
It is shown that the concerns about \textit{understandability, ambiguity, cohesion} of test cases match with the test smells \textit{ambiguous test} and \textit{long/complex/verbose/obscure test}.
Likewise, the concerns about the \textit{complexity} of test cases directly relate to the test smells \textit{eager test}.

\begin{table}
{
    \begin{adjustwidth}{}{}
    \footnotesize
    \begin{center}
    \RaggedRight
    \caption{Test-Case (TC) Quality Classification}
    \label{tab:TCQualityClassification}
    \begin{threeparttable}
    \begin{tabular}{p{0.055\linewidth}p{0.285\linewidth}p{0.215\linewidth}p{0.24\linewidth}p{0.18\linewidth}}
        \toprule
        \textbf{TC ID} & \textbf{Concerns from \mbox{Assessor 1}} & \textbf{Classification (G/B/N)}  & \textbf{Concerns from \mbox{Assessor 2}}  & \textbf{Literature \mbox{\cite{hauptmann2013hunting,garousi2018we}}}\\
        \midrule
        P1.2 & -Understandability: explained objective, links to specs/requirements, unclear precondition \newline -Complexity: combination of multiple TCs \newline -Traceability to bugs: not clear due to the complexity & Assessor1 [P1]: N \newline Assessor2 [P4]: G & -Ambiguity: not well written pre-conditions \newline -Complexity: combination of multiple TCs & -Ambiguous test~\cite{hauptmann2013hunting} \newline -VOLC test~\cite{garousi2018we} \newline -Eager test~\cite{garousi2018we} \\
        \hline
        P4.1 & -Ambiguity: unclear terms, missing expected results of steps, missing pre-conditions & Assessor 1 [P4]: B \newline Assessor 2 [P1]: N	& -Ambiguity: unclear terms \newline -Repeatability: can be run anytime & -Ambiguous test~\cite{hauptmann2013hunting} \newline -VOLC test~\cite{garousi2018we} \\
        \hline
        P5.2 & -Ambiguity: unclear terms \newline -Complexity: combination of multiple TCs \newline -Traceability to bugs: not clear due to the complexity & Assessor 1 [P5]: B \newline Assessor 2 [P2]: N & -Ambiguity: unclear terms & -Ambiguous test~\cite{hauptmann2013hunting} \newline VOLC test~\cite{garousi2018we} \newline -Eager test~\cite{garousi2018we} \\
        \hline 
        P3.1 &  -Understandability: sufficient description & Assessor 1 [P3]: N \newline Assessor 2 [P2]: B & -Ambiguity: unclear terms due to poor English & -Conditional test~\cite{hauptmann2013hunting} \newline -Ambiguous test~\cite{hauptmann2013hunting} \newline -VOLC test~\cite{garousi2018we} \\
        \hline
        \hline
        P3.1 & -Understandability: sufficient description & Assessor1 [P3]: N \newline Assessor2 [P5]: N & -Understandability: explained objective \newline -Ambiguity: missing pre-conditions \newline -Traceability to bugs: established & -Conditional test~\cite{hauptmann2013hunting} \newline -Ambiguous test~\cite{hauptmann2013hunting} \newline -VOLC test~\cite{garousi2018we} \\
        \hline
        P1.3 & -Ambiguity: unclear objective \newline -Complexity: combination of multiple TCs \newline -Traceability to bugs: not clear due to the complexity & Assessor1 [P1]: B \newline Assessor2 [P6]: B & -Ambiguity: unclear objective, unclear terms, unclear expected results for multiple steps \newline -Complexity: combining several TCs & -Conditional test~\cite{hauptmann2013hunting} \newline -Ambiguous test~\cite{hauptmann2013hunting} \newline -VOLC test~\cite{garousi2018we} \newline - Eager test~\cite{garousi2018we} \\
        \hline
        P2.4 & -Ambiguity: missing pre-conditions & Assessor1 [P2]: B \newline Assessor2 [P5]: B & -Ambiguity: unclear objective, missing pre-conditions & -Ambiguous test~\cite{hauptmann2013hunting} \newline -VOLC test~\cite{garousi2018we} \\
        \hline
        P3.2 & -Ambiguity: missing objective \newline -Cohesion: missing steps & Assessor1 [P3]: B \newline Assessor2 [P5]: B &	-Ambiguity: unclear step & Ambiguous test~\cite{hauptmann2013hunting} \newline -VOLC test~\cite{garousi2018we} \\
        \bottomrule
    \end{tabular}
    \begin{tablenotes}
    \footnotesize
    \noindent
    \begin{minipage}[c]{1\linewidth}
    \item[1] VOLC: Long/complex/verbose/obscure~\cite{garousi2018we}
    \end{minipage} 
    \end{tablenotes}
    \end{threeparttable}
    \end{center}
    \end{adjustwidth}
}
\end{table}

However, the concerns about two quality attributes, \textit{traceability} and \textit{repeatability}, have no corresponding smells according to our list of test smells.
One potential reason is that those quality concerns could be the consequences of some other test smells.
\textit{Traceability} could be affected by the test smells \textit{eager test}, \textit{ambiguous test} and \textit{long/complex/verbose/obscure test}.
As pointed out by practitioners, if a test case contains multiple test cases, it becomes complex.
Hence, it is harder to understand which part the test case leads to found issue(s).
Ambiguity in a test case's description could also make the test execution non-deterministic~\cite{hauptmann2013hunting}, which potentially affects the traceability to found issue(s).
Likewise, \textit{repeatability} might not be possible if there are dependencies among the test cases.
Indeed, there are test smells due to dependencies in testing~\cite{garousi2018we}.
However, they were not in our list as they were not the top discussed smells~\cite{garousi2018we}.

\subsection{Quality-related factors (RQ3)}
By answering our interview questions (Q4-9), the practitioners described factors which could influence how they assess test-case quality.
In general, practitioners believe that the test-case quality depends on the test case's context.
For example, the assessment could depend on whether the practitioner knows how the code was written.
He or she might have a different opinion on how to design test cases for testing that code compared with those who do not know the code.
Another context factor is the maturity level of the software system under test (SUT).
According to three practitioners, to save their time, they could combine multiple test cases into one when the SUT is more or less working properly as those test cases hardly fail at that state.
Hence, in that case, a test case is not considered as bad even though it contains different test cases.
Two practitioners mentioned that the testing level also has an impact on how test-case quality is defined.
For example, for exploratory tests, practitioners whose responsible is execute test cases prefer to have flexibility in executing test cases. 
They would rather not to follow steps so closely as that might not help them to identify new issues.
Therefore, if an exploratory test case's execution instructions are restrictive, that test case could be perceived as bad. 
Hence, practitioners' pre-knowledge of the test-case context has a strong influence on their test-case quality perceptions.

\subsection{Improvement (RQ4)}
With the interview question Q14, we identify several suggestions for improving test-case quality.
In general, a homogeneous directive or procedure for test-case design could improve the quality as it could guarantee test cases are designed systematically.
A uniform quality policy could also help to ensure the quality is met and aligned among practitioners.
More specifically, to enhance test-case understandability in the test-case design phase, it was suggested that each test case should contain all necessary information.
Importantly, the information should be relevant to both testers and developers.
That will help to avoid a situation in which testers or developers have to look for information of related test cases in order to understand their assigned test cases.
For test-case maintenance, the most common suggestion was that test cases should be reviewed regularly as they could become obsolete due to the evolution of the SUT.
Updating test cases so that they contain all relevant information for execution and removing no-longer-needed test cases are important steps in this phase.
Apart from improvements in test-case design and maintenance, practitioners also suggested that developers and testers should have active communication in order to mitigate misunderstanding in executing and analysing test results. 

\subsection{Source of Information (RQ5)}
With the interview question Q15, we collected information sources that practitioners refer to for a better understanding of test-case quality.
The most common source is from colleagues like testers and developers working on the same projects, especially seniors who have experience in similar tasks. 
It is consistent with the previous research on information sources consulted by practitioners~\cite{8661213}.

Regarding test-case design, product specifications are considered the most relevant internal source of information.
Other types of internal sources include software architect documents, test cases in previous projects, guidelines and templates for writing test cases, rules and policies from test architects, and test plans.
The practitioners also refer to external sources such as guidelines provided by the ISTQB and ISO standards.
Apart from those common sources, one practitioner also mentioned that he or she learns about test-case quality by attending industrial seminars and workshops on related topics.
Some practitioners also said that they rely on their own experience when assessing test-case quality.

\section{Conclusions and Future Work}\label{sec:conclusion}
We conducted an interview-based exploratory study involving six practitioners, working in three different teams in a company to understand practitioners' perceptions of test-case quality. 
We identified 11 quality attributes for test cases, of which \textit{understandability} was perceived as most important.
That could be due to the nature of the studied test cases, which were written in natural language.
Nevertheless, the study of Garousi et al.~\cite{garousi2018we} also reported the related test smell \textit{long/complex/verbose/obscure test} as the main concern in test code, which means that \textit{understandability} is also important in test code.

We also found that there is a misalignment in practitioners' perceptions of test-case quality.
The explanation is that, depending on the practitioners’ responsibilities, they have different quality requirements. 
For practitioners whose responsibility is to run test cases, the focus is more on acquiring relevant information for test-case execution.
Hence, their priority is the \textit{understandability} of test cases.
For those who need to design and maintain test cases like test architects and developers, their concerns are more about test-case maintenance and outcomes of test suites.
Therefore, they require other quality attributes such as \textit{traceability} to other artefacts, \textit{efficiency, effectiveness, repeatability}, etc.
The context factors of test cases, such as code-related knowledge, the maturity level of software under test, testing types such as exploratory test potentially also impact how practitioners define test-case quality.

We also identified suggestions for improving test-case quality.
The most common suggestion is a homogeneous procedure for test-case design, with focus on completeness of test cases, meaning that a test case should contain all relevant information for execution by any involved party.
Reviewing test cases and regular communication between developers and testers were also highly recommended by practitioners. 
Practitioners also discussed different sources of information they refer for a better understanding of test-case quality.
In general, their information comes from external sources such as ISTQB and ISO standards.
For specific test cases, they rely on the internal sources, such as product specifications, and discussion with other colleagues.

Even though our findings were based on a few data points, we had a sound, repeatable strategy to identify them.
They are not generic, but for a specific context.
For more general findings, we plan to interview more practitioners in different contexts.
We will also compare our findings of the quality attributes and quality definition(s) with other existing studies.
Another planned future work is to have a broader investigation on differences and similarities between the industry and the literature on defining and assessing test-case quality.

\subsubsection{Acknowledgment}
This work has been supported by ELLIIT, a Strategic Area within IT and Mobile Communications, funded by the Swedish Government, and by the VITS project from the Knowledge Foundation Sweden (20180127).

\bibliographystyle{splncs04}
\bibliography{mybib}

\end{document}